\documentclass[twocolumn]{aastex631}

\usepackage{comment}
\usepackage{natbib}
\usepackage{booktabs}
\begin{document}

\title{Impact of Baseline, Cadence, and Host Contamination on AGN Variability Metrics: A Systematic Study with ZTF}

\correspondingauthor{Diego Martínez Collipal}
\email{diemartinez2021@udec.cl}

\author[0009-0004-4993-5656]{Diego Martínez Collipal}
\affiliation{Astronomy Department, Universidad de Concepción, Casilla 160-C, Concepción 4030000, Chile}
\affiliation{Millennium Nucleus on Transversal Research and Technology to Explore Supermassive Black Holes (TITANs)}
\affiliation{Millennium Institute of Astrophysics (MAS), Nuncio Monseñor Sótero Sanz 100, Providencia, Santiago, Chile}

\author[0000-0002-5854-7426]{Swayamtrupta Panda}\thanks{Gemini Science Fellow}
\affiliation{International Gemini Observatory/NSF NOIRLab, Casilla 603, La Serena, Chile}

\begin{abstract}
Variability in active galactic nuclei (AGN) probes the physics of accretion onto supermassive black holes. This variability is characterized using metrics derived from the flux distributions of temporally separated epochs. We studied the stability of two variability metrics, the Stetson index ``\textit{J}" and the smoothness ``\textit{s}", against baseline, cadence, and host galaxy contamination. We studied 23 nearby AGNs using Zwicky Transient Facility's Data Release 24.
Both metrics are robust to baseline variations of $\sim 2$ years. However, \textit{s} is sensitive to cadence, showing variations $\gtrsim 40\%$, while \textit{J} shows minor variations $\lesssim10\%$.
We studied the host galaxy impact using Mrk 493 as a representative case. We found that J remains unchanged after host subtraction, while \textit{s} increases. We concluded that \textit{J} is a robust tool for characterizing AGN variability, while \textit{s} should be interpreted with caution.

\end{abstract}


\section{Introduction} \label{sec:intro}
Active galactic nuclei (AGN) exhibit variability on typical timescales of months to years \citep{Macleod2012}. This variability has multiple possible mechanisms, such as disk instabilities, obscuration, among others. AGN variability is characterized using metrics that quantify the timescales and amplitudes of the episodes. Metrics are also crucial to distinguish AGNs from other variable sources \citep{Butler2011}. \\
AGN metrics are derived from the flux distribution of temporally separated epochs, i.e., light curves, which are constructed from long-term monitoring surveys, such as the Catalina Real-time Transient Survey \citep[CRTS,][]{drake2009}, the Zwicky Transient Facility \citep[ZTF,][]{2019PASP..131a8002B} or the upcoming LSST survey \citep{Ivezi2019}. \\

In this context, we studied two metrics defined in the work \citet{Ma2024} (MA24). The first is the Stetson index ``\textit{J}" \citep{1993AJ....105.1813W}, defined as: 
\begin{equation}
    J = \frac{\sum w_{i}\,sgn (\delta_{i}^{2}-1) \sqrt{|\delta_{i}^{2}-1|}}{\sum w_{i}}
    \label{eq:Jindex}
\end{equation}
where sgn returns the sign of the value, $\delta_{i}$ is defined as 
\begin{equation}
    \delta_{i} = \frac{f_{i}-\bar{f}}{\sigma_{f,i}}\sqrt{\frac{n}{n-1}}
    \label{eq:delta}
\end{equation}
and $w_{i}$ is:
\begin{equation}
    w_{i} = \left[1+\left(\frac{\delta_{i}}{2}\right)^{2}\right]^{-1}
    \label{eq:weight}
\end{equation}
In the equations, $f_{i}$ is the \textit{i}$^{\rm th}$ flux element, $\bar{f}$ the mean flux, $\sigma_{f,i}$ the \textit{i}$^{\rm th}$ uncertainty, and \textit{n} the number of data points. \textit{J} quantify the variability significance against noise. A \textit{J} value $< 0$ indicates that observational noise dominates the variability.\\

Our second metric is the smoothness parameter ``\textit{s}", defined as:
\begin{equation}
    s = \frac{1}{N} \sum \left|\frac{\delta f_{i+1}-\delta f{i}}{t_{i+2}-t_{i}}\right| \frac{\sqrt{\left(f_{i+2}-f_{i+1}\right)^{2} + \left(f_{i+1}-f_{i}\right)^{2}}}{\bar{f}}
    \label{eq:smooth}
\end{equation}
Where \textit{N} is the number of data points and $\delta f_{i}$ is defined as:
\begin{equation}
    \delta f_{i} = \frac{1}{\bar{f}} \frac{f_{i+1}-f_{i}}{t_{i+1}-t_{i}}
    \label{eq:smooth_delta}
\end{equation}

\begin{figure*}[htb!]
\centering
\includegraphics[width=0.925\textwidth]{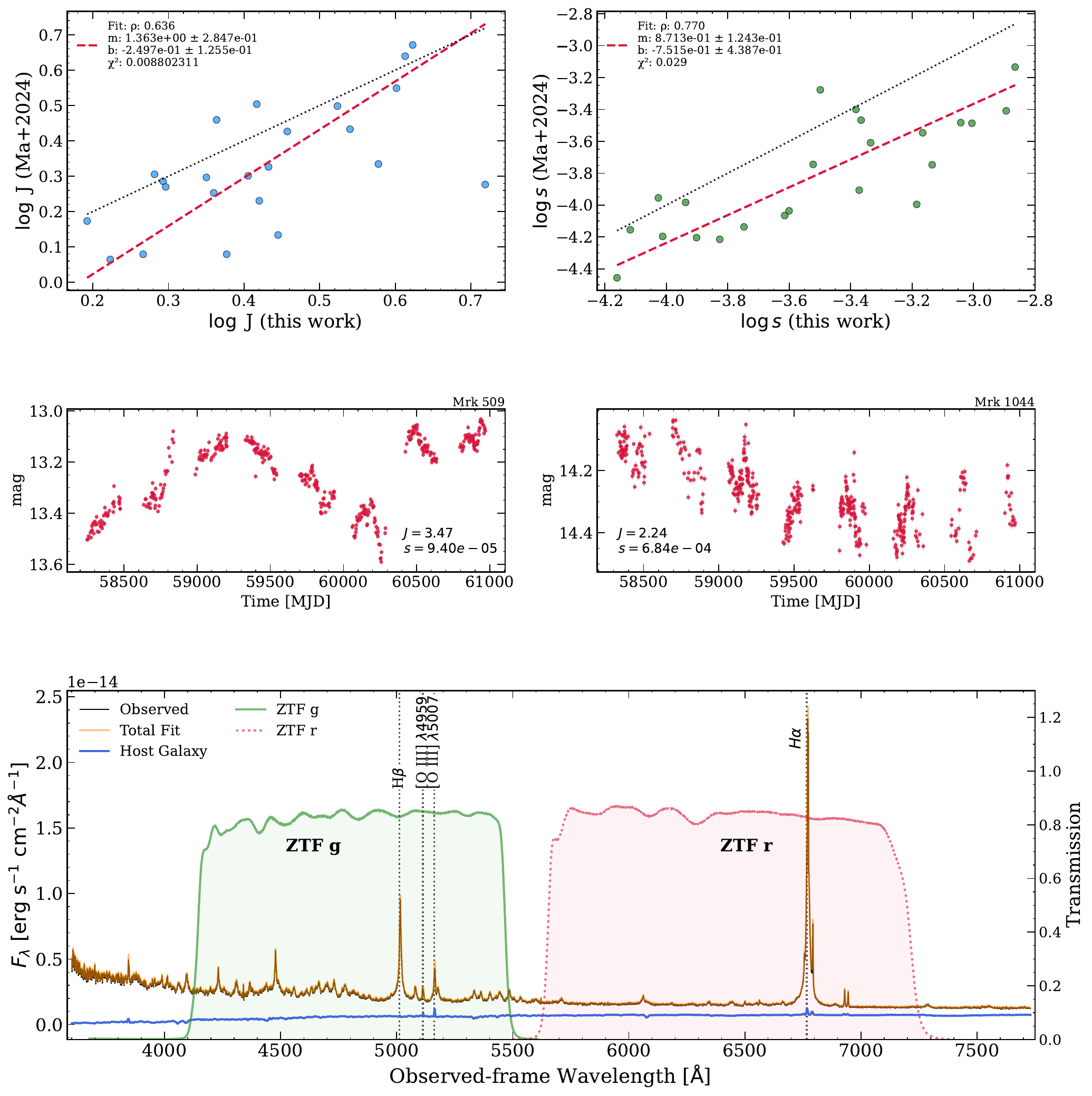}
\caption{\textbf{Upper left}: ZTF DR24 measurements of \textit{J} against those from MA24. The red line shows our fitted linear model obtained via Orthogonal distance regression (ODR), while the black line shows the 1:1 relation. Spearman correlation coefficient $\rho$ and fitted parameters are shown in the top left corner. \textbf{Upper right}: Same as left panel, but for \textit{s}. 
\textbf{Middle panels:} Representative ZTF \textit{r} band light curves. Left: "Assembled" source (high \textit{J}, low \textit{s}). Right: "Dispersed" source (low \textit{J}, high \textit{s}). Each panel shows the source name, \textit{J}, and \textit{s}.   
\textbf{Bottom panel}: Spectral fitting of Mrk 493. The black curve presents the DESI spectrum, while our best {\tt PyQSOfit} is shown in orange. The derived host galaxy contribution is shown in blue. Green (solid) and red (dotted) lines show the transmission curves of ZTF for \textit{g} and \textit{r}, respectively. The transmission values are plotted on a secondary y-axis on the right. We marked the position of the strongest emission lines on each band with vertical dashed lines and text.}
\label{fig:main}
\end{figure*}
\textit{s} evaluates second-order derivatives of the fluxes to quantify the light curve continuity. We associate large \textit{s} values with rapid and erratic fluctuations, which are likely non-physical (MA24). Naturally, we expect that \textit{s} will be affected by the cadence of the observation, penalizing poorly sampled light curves and large observational gaps. Thus, its calculation is restricted to time intervals where $t_{i+2}-t_{i} < 20$ days (MA24). Our equation for \textit{s} includes two modifications with respect to the version of MA24. A normalization factor \textit{N}, and the change in the denominator of the absolute value from $t_{i+1}-t_{i}$ to $t_{i+2}-t_{i}$. These corrections are necessary to replicate the results presented in MA24, and were obtained after private communications with the authors. \\
The present work tested the stability of \textit{J} and \textit{s} against temporal coverage of the observation (baseline), cadence, and host contamination by updating the MA24 measurements using ZTF DR24, which contains data from March 2018 to October 2025. (The original work used DR16, covering data until January 2023.) The comparison between the two sets of light curves from the respective DRs tries to shed some light on the consistency of these metrics over time and determine whether they are comparable between different epochs.  

\section{Sample and data acquisition} \label{sec:analysis}
We studied the sample of 23 AGNs from MA24, which covers redshifts between 0.01 $\lesssim$ z $\lesssim$ 0.09. We retrieved the latest public ZTF DR24 observations \citep{https://doi.org/10.26131/irsa598} for these objects from the IRSA archive\footnote{\url{https://irsa.ipac.caltech.edu/Missions/ztf.html}}. The light curves cover 7+ years of observations, with more than 100 data-points in each band (\textit{g}, \textit{r}, and \textit{i}). \\

We focused on the bands \textit{g} and \textit{r}, as they are better sampled than the \textit{i}-band. We processed the data following MA24. We removed NaN values and 3$\sigma$ outliers from the flux distribution and resampled the light curves into 1-day bins.\\

To study the effect of host contamination, we retrieved the May 2021 (MJD=59358) optical spectrum of one of the sources in our sample (Mrk 493) from DESI DR1 \citep{desi2025}, which falls within our observed ZTF epochs. 

\section{Results and Discussions} \label{sec:results}

\subsection{Baseline effect} \label{subsec:baseline}
We computed \textit{J} and \textit{s} for our ZTF DR24 light curves in the \textit{r}-band\footnote{We provide our Python implementation of \textit{J} and \textit{s} at \url{https://github.com/DiegoMartinez-astro/AGN-variability-techniques}.\dataset[doi: 10.5281/zenodo.19685095]{https://doi.org/10.5281/zenodo.19685095}}, and compared the results with those reported in Table 1 of MA24 (upper panels of Figure \ref{fig:main}). Our estimate of \textit{s} showed agreement with the values of MA24. The measurements are significantly correlated, and our best-fit model is close to the 1:1 relation. For \textit{J}, we found a slightly weaker correlation, but the trend is still linear and close to 1:1 as well. We detected that the object PG 1426+015 significantly deviates from the predicted linear relation (see the rightmost point on the \textit{J} panel). Our \textit{J} value for this object is more than twice that of MA24. This source will be investigated in future work.
\subsection{Cadence effect}
To test cadence impact, we under-sampled our original 1-day binned light curves by dividing each light curve into bins of the desired cadence and randomly selecting a single value per bin. We did this for three bin sizes of 2, 5, and 10 days. For each, we computed \textit{J} and \textit{s}, and compared with the original 1-day case. \\
Taking Mrk 493 in the \textit{r}-band as an example, \textit{J} varies by 3.75\%, 3.22\%, and 11.77\%, for bin sizes of 2,5, and 10 days, respectively, while \textit{s} varies by 76.6\%, 90.10\%, and 98.39\% for the same sizes. This trend holds across our sample, where the average variations for \textit{J} are 4.88\%, 8.55\%, and 16.45\% compared to 59.15\%, 85.80\%, and 96.08\% for \textit{s}. Thus, \textit{J} is robust against cadence, barely exceeding 10\% variations even at 10-day binning, while \textit{s} shows large variations. This is expected, as \textit{s} quantifies the sampling quality of a light curve. Nevertheless, we emphasize that even minor modifications can produce large differences.

\subsection{Host contribution}
We characterize the impact of the host galaxy contamination in \textit{J} and \textbf{s} for Mrk 493. We derived its host spectrum using the spectral-fitting software {\tt PyQSOfit} \citep{pyqsofit}. The fitted model has $\chi_{\nu}^{2} \sim 5.2$ and is presented in Figure \ref{fig:main}. We convolved the host model with the ZTF transmission curves using the package {\tt SNCosmo} \citep{Barbary2016} to obtain \textit{g} and \textit{r} magnitudes. The result revealed that the host accounts for 22\% of the \textit{g}-flux and 34\% in \textit{r}.\\

We subtract this contamination from our ZTF data and recalculated \textit{J} and \textit{s}. After subtracting, \textit{s} increased by factors of 2.33 and 1.64 in the \textit{r} and \textit{g} bands, respectively, while \textit{J} remained unchanged in both. We concluded that the host contamination is significant, especially in the \textit{r} band. This affects the computation of \textit{s}, since host subtraction changes the mean flux in Equation \ref{eq:smooth_delta}, while it does not change \textit{J}, since the host contribution cancels out in the flux differences of Equation \ref{eq:delta}. We emphasize that DESI uses fibers of $\sim 1.5''$ diameter, whereas ZTF integrates flux over a $\sim 5''$ aperture. Thus, ZTF captures more extended emission than DESI and our methodology could underestimate the host contamination.

\begin{acknowledgments}
This research was part of the NOIRLab Chile REU program funded by NSF Award Number 2349023. DMC acknowledges funding from ANID, Chile, through the scholarship ANID MAGISTER NACIONAL, 22251176, and financial support from Millenium Nucleus TITANs (NCN2023${\_}$002, AIM23-0001), and the China-Chile Joint Fund (CCJRF2310). SP is supported by the International Gemini Observatory, a program of NSF NOIRLab, managed by AURA. The authors thank Chi-Zhuo Wang and Xue-Bing Wu for helpful discussions regarding the smoothness parameter.
\end{acknowledgments}

\bibliography{references}{}

@ARTICLE{Ma2024,
       author = {{Ma}, Qinchun and {Wen}, Yuhan and {Wu}, Xue-Bing and {Gu}, Huapeng and {Fu}, Yuming},
        title = "{H{\ensuremath{\alpha}} Time Delays of Active Galactic Nuclei from the Zwicky Transient Facility Broadband Photometry}",
      journal = {\apj},
         year = 2024,
        month = may,
       volume = {966},
       number = {1},
          eid = {5},
        pages = {5},
          doi = {10.3847/1538-4357/ad34d6},
archivePrefix = {arXiv},
       eprint = {2403.10223},
 primaryClass = {astro-ph.GA},
       adsurl = {https://ui.adsabs.harvard.edu/abs/2024ApJ...966....5M}
}

@ARTICLE{Macleod2012,
       author = {{MacLeod}, Chelsea L. and {Ivezi{\'c}}, {\v{Z}}eljko and {Sesar}, Branimir and {de Vries}, Wim and {Kochanek}, Christopher S. and {Kelly}, Brandon C. and {Becker}, Andrew C. and {Lupton}, Robert H. and {Hall}, Patrick B. and {Richards}, Gordon T. and {Anderson}, Scott F. and {Schneider}, Donald P.},
        title = "{A Description of Quasar Variability Measured Using Repeated SDSS and POSS Imaging}",
      journal = {\apj},
         year = 2012,
        month = jul,
       volume = {753},
       number = {2},
          eid = {106},
        pages = {106},
          doi = {10.1088/0004-637X/753/2/106},
archivePrefix = {arXiv},
       eprint = {1112.0679},
 primaryClass = {astro-ph.CO},
       adsurl = {https://ui.adsabs.harvard.edu/abs/2012ApJ...753..106M}
}

@ARTICLE{Butler2011,
       author = {{Butler}, Nathaniel R. and {Bloom}, Joshua S.},
        title = "{Optimal Time-series Selection of Quasars}",
      journal = {\aj},
         year = 2011,
        month = mar,
       volume = {141},
       number = {3},
          eid = {93},
        pages = {93},
          doi = {10.1088/0004-6256/141/3/93},
archivePrefix = {arXiv},
       eprint = {1008.3143},
 primaryClass = {astro-ph.CO},
       adsurl = {https://ui.adsabs.harvard.edu/abs/2011AJ....141...93B}
}

@ARTICLE{Ivezi2019,
       author = {{Ivezi{\'c}}, {\v{Z}}eljko and {Kahn}, Steven M. and {Tyson}, J. Anthony and {Abel}, Bob and {Acosta}, Emily and {Allsman}, Robyn and {Alonso}, David and {AlSayyad}, Yusra and {Anderson}, Scott F. and {Andrew}, John and {Angel}, James Roger P. and {Angeli}, George Z. and {Ansari}, Reza and {Antilogus}, Pierre and {Araujo}, Constanza and {Armstrong}, Robert and {Arndt}, Kirk T. and {Astier}, Pierre and {Aubourg}, {\'E}ric and {Auza}, Nicole and {Axelrod}, Tim S. and {Bard}, Deborah J. and {Barr}, Jeff D. and {Barrau}, Aurelian and {Bartlett}, James G. and {Bauer}, Amanda E. and {Bauman}, Brian J. and {Baumont}, Sylvain and {Bechtol}, Ellen and {Bechtol}, Keith and {Becker}, Andrew C. and {Becla}, Jacek and {Beldica}, Cristina and {Bellavia}, Steve and {Bianco}, Federica B. and {Biswas}, Rahul and {Blanc}, Guillaume and {Blazek}, Jonathan and {Blandford}, Roger D. and {Bloom}, Josh S. and {Bogart}, Joanne and {Bond}, Tim W. and {Booth}, Michael T. and {Borgland}, Anders W. and {Borne}, Kirk and {Bosch}, James F. and {Boutigny}, Dominique and {Brackett}, Craig A. and {Bradshaw}, Andrew and {Brandt}, William Nielsen and {Brown}, Michael E. and {Bullock}, James S. and {Burchat}, Patricia and {Burke}, David L. and {Cagnoli}, Gianpietro and {Calabrese}, Daniel and {Callahan}, Shawn and {Callen}, Alice L. and {Carlin}, Jeffrey L. and {Carlson}, Erin L. and {Chandrasekharan}, Srinivasan and {Charles-Emerson}, Glenaver and {Chesley}, Steve and {Cheu}, Elliott C. and {Chiang}, Hsin-Fang and {Chiang}, James and {Chirino}, Carol and {Chow}, Derek and {Ciardi}, David R. and {Claver}, Charles F. and {Cohen-Tanugi}, Johann and {Cockrum}, Joseph J. and {Coles}, Rebecca and {Connolly}, Andrew J. and {Cook}, Kem H. and {Cooray}, Asantha and {Covey}, Kevin R. and {Cribbs}, Chris and {Cui}, Wei and {Cutri}, Roc and {Daly}, Philip N. and {Daniel}, Scott F. and {Daruich}, Felipe and {Daubard}, Guillaume and {Daues}, Greg and {Dawson}, William and {Delgado}, Francisco and {Dellapenna}, Alfred and {de Peyster}, Robert and {de Val-Borro}, Miguel and {Digel}, Seth W. and {Doherty}, Peter and {Dubois}, Richard and {Dubois-Felsmann}, Gregory P. and {Durech}, Josef and {Economou}, Frossie and {Eifler}, Tim and {Eracleous}, Michael and {Emmons}, Benjamin L. and {Fausti Neto}, Angelo and {Ferguson}, Henry and {Figueroa}, Enrique and {Fisher-Levine}, Merlin and {Focke}, Warren and {Foss}, Michael D. and {Frank}, James and {Freemon}, Michael D. and {Gangler}, Emmanuel and {Gawiser}, Eric and {Geary}, John C. and {Gee}, Perry and {Geha}, Marla and {Gessner}, Charles J.~B. and {Gibson}, Robert R. and {Gilmore}, D. Kirk and {Glanzman}, Thomas and {Glick}, William and {Goldina}, Tatiana and {Goldstein}, Daniel A. and {Goodenow}, Iain and {Graham}, Melissa L. and {Gressler}, William J. and {Gris}, Philippe and {Guy}, Leanne P. and {Guyonnet}, Augustin and {Haller}, Gunther and {Harris}, Ron and {Hascall}, Patrick A. and {Haupt}, Justine and {Hernandez}, Fabio and {Herrmann}, Sven and {Hileman}, Edward and {Hoblitt}, Joshua and {Hodgson}, John A. and {Hogan}, Craig and {Howard}, James D. and {Huang}, Dajun and {Huffer}, Michael E. and {Ingraham}, Patrick and {Innes}, Walter R. and {Jacoby}, Suzanne H. and {Jain}, Bhuvnesh and {Jammes}, Fabrice and {Jee}, M. James and {Jenness}, Tim and {Jernigan}, Garrett and {Jevremovi{\'c}}, Darko and {Johns}, Kenneth and {Johnson}, Anthony S. and {Johnson}, Margaret W.~G. and {Jones}, R. Lynne and {Juramy-Gilles}, Claire and {Juri{\'c}}, Mario and {Kalirai}, Jason S. and {Kallivayalil}, Nitya J. and {Kalmbach}, Bryce and {Kantor}, Jeffrey P. and {Karst}, Pierre and {Kasliwal}, Mansi M. and {Kelly}, Heather and {Kessler}, Richard and {Kinnison}, Veronica and {Kirkby}, David and {Knox}, Lloyd and {Kotov}, Ivan V. and {Krabbendam}, Victor L. and {Krughoff}, K. Simon and {Kub{\'a}nek}, Petr and {Kuczewski}, John and {Kulkarni}, Shri and {Ku}, John and {Kurita}, Nadine R. and {Lage}, Craig S. and {Lambert}, Ron and {Lange}, Travis and {Langton}, J. Brian and {Le Guillou}, Laurent and {Levine}, Deborah and {Liang}, Ming and {Lim}, Kian-Tat and {Lintott}, Chris J. and {Long}, Kevin E. and {Lopez}, Margaux and {Lotz}, Paul J. and {Lupton}, Robert H. and {Lust}, Nate B. and {MacArthur}, Lauren A. and {Mahabal}, Ashish and {Mandelbaum}, Rachel and {Markiewicz}, Thomas W. and {Marsh}, Darren S. and {Marshall}, Philip J. and {Marshall}, Stuart and {May}, Morgan and {McKercher}, Robert and {McQueen}, Michelle and {Meyers}, Joshua and {Migliore}, Myriam and {Miller}, Michelle and {Mills}, David J.},
        title = "{LSST: From Science Drivers to Reference Design and Anticipated Data Products}",
      journal = {\apj},
         year = 2019,
        month = mar,
       volume = {873},
       number = {2},
          eid = {111},
        pages = {111},
          doi = {10.3847/1538-4357/ab042c},
archivePrefix = {arXiv},
       eprint = {0805.2366},
 primaryClass = {astro-ph},
       adsurl = {https://ui.adsabs.harvard.edu/abs/2019ApJ...873..111I}
}

@ARTICLE{desi2025,
       author = {{DESI Collaboration} and {Karim}, M. Abdul and {Adame}, A.~G. and {Aguado}, D. and {Aguilar}, J. and {Ahlen}, S. and {Alam}, S. and {Aldering}, G. and {Alexander}, D.~M. and {Alfarsy}, R. and {Allen}, L. and {Allende Prieto}, C. and {Alves}, O. and {Anand}, A. and {Andrade}, U. and {Armengaud}, E. and {Avila}, S. and {Aviles}, A. and {Awan}, H. and {Bailey}, S. and {Baleato Lizancos}, A. and {Ballester}, O. and {Bault}, A. and {Bautista}, J. and {Bean}, R. and {Behera}, J. and {BenZvi}, S. and {Beraldo e Silva}, L. and {Bermejo-Climent}, J.~R. and {Beutler}, F. and {Bianchi}, D. and {Blake}, C. and {Blum}, R. and {Bolton}, A.~S. and {Bonici}, M. and {Brieden}, S. and {Brodzeller}, A. and {Brooks}, D. and {Buckley-Geer}, E. and {Burtin}, E. and {Bystr{\"o}m}, A. and {Canning}, R. and {Carnero Rosell}, A. and {Carr}, A. and {Carrilho}, P. and {Casas}, L. and {Castander}, F.~J. and {Cereskaite}, R. and {Cervantes-Cota}, J.~L. and {Chaussidon}, E. and {Chaves-Montero}, J. and {Chen}, S. and {Chen}, X. and {Circosta}, C. and {Claybaugh}, T. and {Cole}, S. and {Cooper}, A.~P. and {Cousinou}, M.-C. and {Cuceu}, A. and {Davis}, T.~M. and {Dawson}, K.~S. and {de Belsunce}, R. and {de la Cruz}, R. and {de la Macorra}, A. and {de Mattia}, A. and {Deiosso}, N. and {Della Costa}, J. and {Demina}, R. and {Demirbozan}, U. and {DeRose}, J. and {Dey}, A. and {Dey}, B. and {Ding}, J. and {Ding}, Z. and {Doel}, P. and {Douglass}, K. and {Dowicz}, M. and {Ebina}, H. and {Edelstein}, J. and {Eisenstein}, D.~J. and {Elbers}, W. and {Emas}, N. and {Escoffier}, S. and {Fagrelius}, P. and {Fan}, X. and {Fanning}, K. and {Favole}, G. and {Fawcett}, V.~A. and {Fern{\'a}ndez-Garc{\'\i}a}, E. and {Ferraro}, S. and {Findlay}, N. and {Font-Ribera}, A. and {Forero-Romero}, J.~E. and {Forero-S{\'a}nchez}, D. and {Frenk}, C.~S. and {G{\"a}nsicke}, B.~T. and {Galbany}, L. and {Garc{\'\i}a-Bellido}, J. and {Garcia-Quintero}, C. and {Garrison}, L.~H. and {Gazta{\~n}aga}, E. and {Gil-Mar{\'\i}n}, H. and {Gloudemans}, A. and {Gnedin}, O.~Y. and {Gontcho}, S. Gontcho A and {Gonzalez}, D. and {Gonzalez-Morales}, A.~X. and {Gonzalez-Perez}, V. and {Gordon}, C. and {Graur}, O. and {Green}, D. and {Gruen}, D. and {Gsponer}, R. and {Guandalin}, C. and {Gutierrez}, G. and {Guy}, J. and {Hahn}, C. and {Han}, J.~J. and {Han}, J. and {He}, S. and {Herrera-Alcantar}, H.~K. and {Heydenreich}, S. and {Honscheid}, K. and {Hou}, J. and {Howlett}, C. and {Huterer}, D. and {Ir{\v{s}}i{\v{c}}}, V. and {Ishak}, M. and {Jacques}, A. and {Jiang}, L. and {Jimenez}, J. and {Jing}, Y.~P. and {Joachimi}, B. and {Joudaki}, S. and {Joyce}, R. and {Jullo}, E. and {Juneau}, S. and {Kara{\c{c}}ayl{\i}}, N.~G. and {Karim}, T. and {Kehoe}, R. and {Kent}, S. and {Khederlarian}, A. and {Kirkby}, D. and {Kisner}, T. and {Kitaura}, F.-S. and {Kizhuprakkat}, N. and {Kong}, H. and {Koposov}, S.~E. and {Kremin}, A. and {Krolewski}, A. and {Lahav}, O. and {Lai}, Y. and {Lamman}, C. and {Lan}, T.-W. and {Landriau}, M. and {Lang}, D. and {Lange}, J.~U. and {Lasker}, J. and {Le Goff}, J.~M. and {Le Guillou}, L. and {Leauthaud}, A. and {Levi}, M.~E. and {Li}, S. and {Li}, T.~S. and {Liu}, W. and {Lodha}, K. and {Lokken}, M. and {Luo}, Y. and {Luo}, Y. and {Magneville}, C. and {Manera}, M. and {Manser}, C.~J. and {Margala}, D. and {Martini}, P. and {Maus}, M. and {McCullough}, J. and {McDonald}, P. and {Medina}, G.~E. and {Medina-Varela}, L. and {Meisner}, A. and {Mena-Fern{\'a}ndez}, J. and {Menegas}, A. and {Meneses-Rizo}, J. and {Mezcua}, M. and {Miquel}, R. and {Montero-Camacho}, P. and {Moon}, J. and {Moustakas}, J. and {Mu{\~n}oz-Guti{\'e}rrez}, A. and {Mu{\~n}oz-Santos}, D. and {Myers}, A.~D. and {Myles}, J. and {Nadathur}, S. and {Najita}, J. and {Napolitano}, L. and {Newman}, J.~A. and {Nikakhtar}, F. and {Nikutta}, R. and {Niz}, G. and {Noriega}, H.~E.},
        title = "{Data Release 1 of the Dark Energy Spectroscopic Instrument}",
      journal = {arXiv e-prints},
         year = 2025,
        month = mar,
          eid = {arXiv:2503.14745},
        pages = {arXiv:2503.14745},
          doi = {10.48550/arXiv.2503.14745},
archivePrefix = {arXiv},
       eprint = {2503.14745},
 primaryClass = {astro-ph.CO},
       adsurl = {https://ui.adsabs.harvard.edu/abs/2025arXiv250314745D}
}

@software{pyqsofit,
       author = {{Guo}, Hengxiao and {Shen}, Yue and {Wang}, Shu},
        title = "{PyQSOFit: Python code to fit the spectrum of quasars}",
         howpublished = {Astrophysics Source Code Library, record ascl:1809.008},
         year = 2018,
        month = sep,
          eid = {ascl:1809.008},
archivePrefix = {ascl},
       eprint = {1809.008},
       adsurl = {https://ui.adsabs.harvard.edu/abs/2018ascl.soft09008G}
}

@misc{https://doi.org/10.26131/irsa598,
  doi = {10.26131/IRSA598},
  url = {https://catcopy.ipac.caltech.edu/dois/doi.php?id=10.26131/IRSA598},
  author = {{ZTF Team}},
  title = {ZTF Lightcurves},
  publisher = {IPAC},
  year = {2025}
}
\bibliographystyle{aasjournal}


\end{document}